# A novel small-scale self-focusing suppression method for ultrahigh peak power lasers

Shuren Pan,[1,2,5] Fenxiang Wu,[1,5] Jiabing Hu,[1] Yang Zhao,[1] Zongxin Zhang,[1] Yi Xu,[1,4,6] Yuxin Leng,[1,4,7] Ruxin Li,[1] Efim Khazanov[3,4]

[1]State Key Laboratory of High Field Laser Physics and CAS Center for Excellence in Ultra-intense Laser Science, Shanghai Institute of Optics and Fine Mechanics (SIOM), Chinese Academy of Sciences, Shanghai 201800, China
[2]Center of Materials Science and Optoelectronics Engineering, University of Chinese Academy of Sciences, Beijing 100049, China
[3]Institute of Applied Physics (IAP) of Russian Academy of Science
[4] IAP-SIOM Joint Laser Laboratory
[5]These authors contribute equally to this work
[6]xuyi@siom.ac.cn
[7]lengyuxin@mail.siom.ac.cn

**A novel method, combining an asymmetric four-grating compressor (AFGC) with the pulse post-compression, is proposed to improve the spatial uniformity of laser beams and hence to suppress the small-scale self-focusing (SSSF) effect during the interaction between ultrahigh peak power lasers and nonlinear materials. The spatial uniformity of laser beams is an important factor in performing post-compression, due to the spatial intensity modulation will be enhanced during the nonlinear propagation process and then seriously damage following optical components. Three-dimensional numerical simulations of the post-compression are implemented based on a petawatt (PW) class laser with a standard compressor and an AFGC respectively. The results indicate that the post-compression with AFGC can efficiently suppress the SSSF and meanwhile shorten the laser pulses from ~30fs to sub-10fs. This work can provide a promising route to overcome the challenge of SSSF and will be very meaningful to promote the practical application of post-compression in ultrahigh peak power lasers.**

Thanks to the developments of chirped pulse amplification (CPA) and optical parametric chirped pulse amplification (OPCPA) [1-2] techniques, the laser peak power has reached 10 PW level, and the corresponding laser focused intensity has reached $10^{22}$ W/cm$^2$ and even$10^{23}$ W/cm$^2$ [3-7]. Such super-intense lasers can bring several significant breakthrough and advance for high-field sciences, and open fascinating and entirely new applications [8-10]. Nowadays, some countries have also commissioned the construction of 100 PW level lasers [11-14], to pursue higher laser intensity and thereby to explore frontier sciences. As is well known, the laser peak power can be enhanced by increasing energy or by further shortening the pulse duration after the laser compressor. However, the limitation of peak power enhancement is no longer coming from the pulse amplification, but from the pulse compression which is restricted by the available size and damage threshold of compression gratings. And a multifold peak power enhancement can only be realized by using mosaic gratings in laser compressor or coherent combination with several grating compressors. As a result, it entails a significant increase in the complexity, size and cost of ultrahigh peak power lasers. Hence, the pulse duration shortening after laser compressor, i.e. post-compression, is obviously a promising approach to further enhance the peak power of lasers which can be carried out without adding costly amplifiers and compression gratings. In addition, it is notable that the post-compression technique is also a potential way to develop compact and economical high peak power lasers.

The post-compression process generally consists of the spectral broadening by self-phase modulation and the pulse recompression by compensation of the newly introduced dispersion with chirped mirrors [15-16]. For the post-compression of high peak power lasers, the spectral broadening based on solid thin plates has been demonstrated to be a proper way to shorten the pulse duration. In 2018, S. K. Lee *et al.* reported a post-compression of a high energy laser from 31fs / 113TW to 14.5fs / 213TW, by using a 500um fused silica plate [17]. In 2020, V. Ginzburg *et al.* demonstrated a fivefold shortening (from 70fs to 14fs) of the 250TW-level PEARL laser, by using three 1mm-thick fused silica plates [18]. In 2021, they further compressed the laser pulses to 11fs through using a 5mm-thick silica plate [19]. In addition to silica plates, the PEARL laser has also been compressed to 10fs by using a 4mm-thick KDP (potassium dihydrogen phosphate) crystal in 2021 [20]. Recently, JI IN KIM *et al.* have also proved the generation of sub-10fs pulses through the post-compression of a 100TW class laser using a 1.5mm-thick fused silica plate, and a spatial filter is installed before laser compressor to suppress the SSSF induced intensity spikes [21]. Moreover, as early as 2013 Aleksandr A. Voronin *et al.* proposed and simulated the subexawatt few-cycle pulse generation via the post-compression of a 13PW / 120fs laser [22].

Although the post-compression technique features undisputed merits, there are still some problems hindering its implementation, especially the SSSF. The SSSF mainly arises from the nonuniformity of laser beam intensity, and is presented even in a plane wave, as it is induced by Bespalov-Talanov instability [23]. The SSSF can result in a significant impairment of beam quality, uncontrollable spectral broadening, and the breakdown of optical components. As is well known, the spatial intensity and phase modulation of high peak power laser beam is usually relatively high, which is mainly induced by the defect of high-energy pump lasers and large-size gain media. Besides, hot spots will also appear due to the diffraction by dusts or defects on optics. Due to the spatial intensity modulation or the hot spots will be further aggravated during the nonlinear propagation process. And thus the SSSF suppression has become a key issue of the post-compression in high peak power lasers. On the one hand, the SSSF can be suppressed by filtering the spatial perturbations during beam propagation in free space, i.e. beam self-filtering with a specific spatial distance, which has been demonstrated to be an efficient suppression method [17-25]. On the other hand, the impact of SSSF can be also suppressed by improving the spatial uniformity of the laser beams before post-compression. To this end, the spatial filters were employed before laser compressor [21-22]. However, spatial filters can only filter the high-frequency modulations of laser beam, and the effect of beam smoothing is very finite.

In this work, a novel method is proposed and demonstrated for suppressing the SSSF during post-compression in high peak power lasers. The core of this method is to improve the spatial intensity uniformity by using an AFGC before post-compression. It provides a promising route to overcome the challenge of SSSF in traditional post-compression. Three-dimensional numerical simulations of the post-compression are implemented based on a PW-class laser with a standard compressor and an AFGC. The numerical results show that the post-compression with AFGC can effectively suppress the SSSF effect and meanwhile shorten the laser pulses from ~30fs to sub-10fs. This work should be meaningful to promote the practical application of post-compression in ultrahigh peak power lasers.

A typical post-compression scheme of high peak power lasers is shown in Fig.1. The output compressed laser beam is firstly spectral broadened by nonlinear transmission through thin plates (TTPs), and then dispersion compensated by chirped mirrors (CMs).

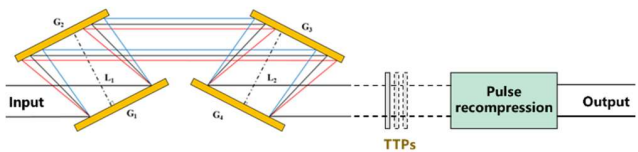

Fig. 1. Setup of traditional post-compression of high peak power lasers.

A three-dimensional simulation of above scheme is implemented based on a 1PW laser. Its output compressed laser beam diameter is ~210mm, pulse spectrum is centered at 800nm with ~60nm full width at half maximum (at FWHM), pulse duration is ~30fs, peak power is ~1PW, corresponding to a laser intensity of ~2.9TW/cm². Figure 2(a) shows a simulated spatial beam profile of this laser, and the PTA (peak-to-average) of spatial intensity distribution is 2.92.

The intensity modulation curves correspond to the X- and Y-axis at the center of beam profile, and the same goes for the following beam profiles. The pixel size of these laser beam profiles is about 300μm. The nonlinear medium for spectral broadening are fused silica TTPs, with the group velocity dispersion of ~36fs²/mm and the nonlinear refractive index of 2.4×10⁻⁷ cm²/GW at 800nm central wavelength. The laser propagation in TTPs is simulated by a three-dimensional time-dependent generalized nonlinear Schrödinger equation [26], shown as Eq.1.

$$\frac{\partial A}{\partial z} = -\frac{\alpha}{2}A - \left(\sum_{k\geq 2}\beta_n \frac{i^{n-1}}{n!}\frac{\partial^n}{\partial T^n}\right)A + i\gamma\left(1 + \frac{1}{\omega_0}\frac{\partial}{\partial T}\right)$$
$$\times \left((1-f_R)A|A|^2 + f_R A \int_0^\infty h_R(\tau)|A(z,T-\tau)|^2 d\tau\right) \quad (1)$$

The attainable recompressed pulse durations of above PW laser beam are calculated based on the different length of TTPs. As shown in Fig.2(b), sub-10fs recompressed laser pulses are achievable in the case of 1.5mm-thick fused silica TTPs. Furthermore, the post-compression based on 1.5mm-thick fused silica thin plates has also been experimentally demonstrated in a similar high peak power laser [21]. Hence, the following simulations will be performed based on 1.5mm-thick fused silica plates.

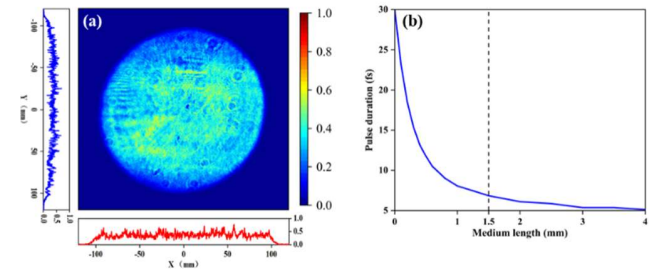

Fig. 2. (a) The beam profile of a PW-class laser, (b) the relation between the achievable recompression pulse durations and the length of TTPs.

After passing through above 1.5mm-thick fused silica TTPs, the spatial beam profile of spectral broadened pulse is shown as Fig.3. Obviously, serious SSSF occurs during above spectral broadening process, and several strong intensity spikes are produced. Such intensity spikes may significantly damage the subsequently optical elements and can prevent the practical application of post-compression. Therefore, the spatial beam quality of is a core factor in performing post-compression of ultrahigh peak power lasers.

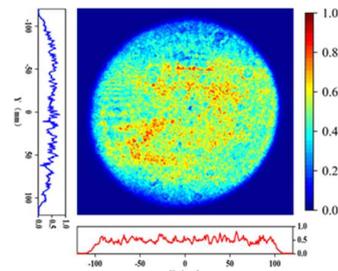

Fig. 3. The spatial beam profile of spectral broadened laser, with a PTA of 2.15.

To overcome the challenge of SSSF, an AFGC is carried out before the post-compression, which can greatly smooth the spatial profile by introducing a moderate spatial chirp across laser beam [27]. The optical configuration of an AFGC is shown as Fig.4. Different from a standard four-grating compressor with a symmetric structure, the two grating pair separations of AFGC are unequal, i.e. $L_1 \neq L_2$. In such case, some spatial chirp will be introduced to the output laser beam from the compressor, and the spatial chirp length $d_0$ can be written as Eq.2. Where, $\omega_l$ and $\omega_s$ represents the shortest and the longest wavelength component of the input laser respectively, and $\alpha$ is the incident angle on $G1$. Clearly, spatial chirp length $d_0$ is proportional to the difference between the two grating pair separations, i.e. ($L_2-L_1$). Moreover, owning to the relatively strong diffraction ability of gratings, large spatial chirp can be realized just with a small different between $L_1$ and $L_2$.

$$d_0 = (\tan\beta(\omega_s) - \tan\beta(\omega_l)) \times \cos\alpha \times (L_2 - L_1). \quad (2)$$

It is worth noting that an AFGC can also provide absolute the same amount of temporal chirp to compress the amplified pulses in comparison to a standard four grating compressor, as long as ($L_2+L_1$) are equal in both designs. Thereby, the impact of residual temporal chirp on the following spectral broadening process can be avoided in the case of AFGC. Moreover, an AFGC with moderate spatial chirp barely brings a little impact to the far-field beam spot, which has been demonstrated by our previous studies [28-29].

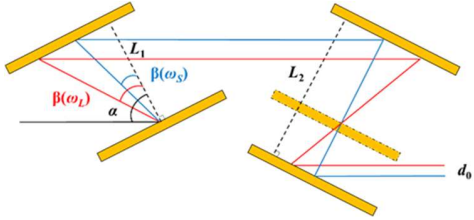

Fig. 4. The optical configuration of an AFGC.

To demonstrate the advantages of AFGC in SSSF suppression, a numerical simulation is also carried out based on above PW laser. Firstly, the compressor is adjusted to be an AFGC for improving the spatial uniformity of laser beam. To this end, the separations of two grating pairs are altered respectively, while keeping the sum of two separations constant. The beam smoothing of an AFGC is simulated based on Reference 27. The relation between the spatial dispersion length of above AFGC and the PTA of the spatial beam profile after AFGC is shown as Fig.5(a). It can be found that the PTA is decreasing with spatial dispersion length increases, and the beam smoothing effect induced by the first 10mm spatial dispersion length is most significant.

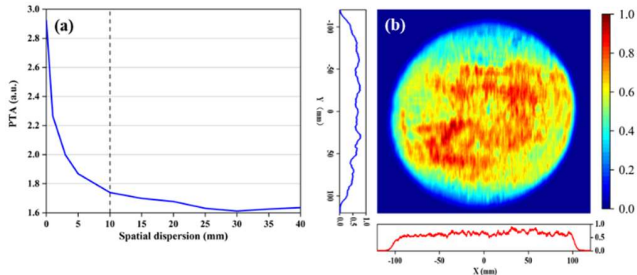

Fig. 5. (a) the relation between the space dispersion length of AFGC and the PTA of spatial beam profile after AFGC, (b) the output spatial beam profile after AFGC with a spatial dispersion length of 10mm.

In following simulations, a 10mm spatial dispersion length is introduced by the AFGC, the separations of each grating pair are 1.068m and 0.932m respectively. The laser incident angle of the laser compressor is 50°, the grating groove density is 1480gr/mm. Comparing with large beam size of 200mm, the small spatial dispersion length (~10mm) would just cause a slight impact to the far-field beam spot, which is ignored in this work. The near-field beam profile after this AFGC is shown as Fig.5(b). Obviously, the spatial beam quality is significantly improved, and the PTA of the spatial intensity is reduced from 2.92 to 1.56.

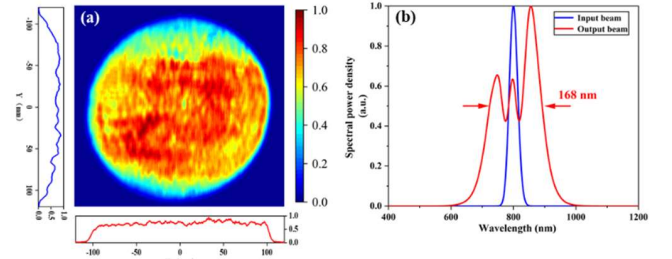

Fig. 6. (a) The spatial beam profiles and (b) spectrum of the laser beam after spectral broadening stage.

Then, the smooth laser beam passes through 1.5mm-thick fused silica TTPs for spectral broadening, as the same as the case with a standard laser compressor. The spatial beam profile after spectral broadening stage is shown in Fig.6(a). Compared with the result in Fig.3, the SSSF is effectively suppressed, and the PTA of spatial intensity distribution is greatly decreased from 2.15 to 1.54. The output spectrum is shown in Fig.6(b), with a bandwidth of 168nm at FWHM. The asymmetry of broadened spectrum is mainly induced by the self-steepening effect.

Lastly, the residual dispersion of spectral broadened laser beam will be compensated by some large-aperture broadband CMs. The additional GDD caused by SPM process and the material dispersion of nonlinear materials, is about 80fs$^2$. After being reflected by two broadband CMs with a total GDD of -80fs$^2$, sub-10fs ultrashort laser pulses are expected, as shown in Fig.7.

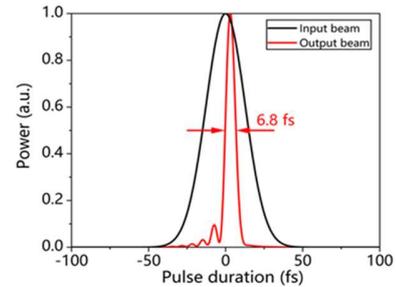

Fig.7. The pulse durations before and after post-compression with an AFGC.

In conclusion, a novel method is proposed for the first time to suppress the SSSF during the post-compression of ultrahigh peak power lasers. The key of this method is the effective combination of an AFGC with post-compression, in which case the beam smoothing of AFGC can be fully utilized in the following post-compression. The numerical results show that the SSSF can be effectively suppressed by this method. Hence, this method should be meaningful to further promote the practical application of post-compression in ultrahigh peak power lasers. Limited by our own computing resources, some details on spatial beam profile improvement may not be revealed in this work, and the more detailed investigations with a much higher spatial resolution will be carried out in the following work.

**Funding.** National Key R&D Program of China (2022YFE0204800, 2022YFA1604401, 2019YFF01014401); Shanghai Sailing Program (21YF1453800); The National Natural Science Foundation of China (11127901, 61925507); The International Partnership Program of Chinese Academy of Sciences (181231KYSB20200040); Shanghai Science and Technology Committee Program (22560780100, 23560750200); Chinese Academy of Sciences President's International Fellowship Initiative (2023VMB0008); Youth Innovation Promotion Association of the Chinese Academy of Sciences.

**Disclosures.** The authors declare no conflicts of interest.

**Data availability.** The data are available from the authors upon reasonable request.

## References

1. M. D. Strickland, G. Mourou, Opt. Commun. **55**, 219 (1985).
2. Dubietis, G. Jonušauskas, A. Piskarskas, Opt. Commun. **88**, 437 (1992).
3. Z. Zhang, F. Wu, J. Hu, X. Yang, J. Gui, X. Liu, C. Wang, Y. Liu, X. Lu, Y. Xu, Y. Leng, R. Li, and Z. Xu, High Power Laser Sci. Eng. **8**, e4 (2020).
4. H. Kiriyama, Y. Miyasak, A. Kon, M. Nishiuchi, A. Sagisaka, H. Sasao, A. S. Pirozhkov, Y. Fukuda, K. Ogura, K. Kondo, N. P. Dover, M. Kando, High Power Laser Sci. Eng. **9**, e62 (2021).
5. F. Lureau, G. Matras, O. Chalus, C. Derycke, T. Morbieu, C. Radier, O. Casagrande, S. Laux, S. Ricaud, G. Rey, A. Pellegrina, C. Richard, L. Boudjemaa, C. S. Boisson, A. Baleanu, R. Banici, A. Gradinariu, C. Caldararu, B. D. Boisdeffre, P. Ghenuche, A. Naziru, G. Kolliopoulos, L. Neagu, R. Dabu, I. Dancus, D. Ursescu, High Power Laser Sci. Eng. **8**, e43 (2020).
6. S. Borneis, T. Laštovička, M. Sokol, T.-M. Jeong, F. Condamine, O. Renner, V. Tikhonchuk, H. Bohlin, A. Fajstavr, J.-C. Hernandez, N. Jourdain, D. Kumar, D. Modřanský, A. Pokorný, A. Wolf, S. Zhai, G. Korn and S. Weber, High Power Laser Sci. Eng. **9**, e30 (2021).
7. J. W. Yoon, Y. G. Kim, I. W. Choi, J. H. Sung, H. W. Lee, S. K. Lee, C. H. Nam, Optica **8**(5), 630 (2021).
8. D. Umstadter, Phys. Plasmas **8**, 1774 (2001).
9. A. J. Gonsalves, K. Nakamura, J. Daniels, C. Benedetti, C. Pieronek, T. C. H. de Raadt, S. Steinke, J. H. Bin, S. S. Bulanov, J. van Tilborg, C. G. R. Geddes, C. B. Schroeder, Cs. Tóth, E. Esarey, K. Swanson, L. Fan-Chiang, G. Bagdasarov, N. Bobrova, V. Gasilov, G. Korn, P. Sasorov, W. P. Leemans, Phys. Rev. Lett. **122**, 084801 (2019).
10. W. Wang, K. Feng, L. Ke, C. Yu, Y. Xu, R. Qi, Y. Chen, Z. Qin, Z. Zhang, M. Fang, J. Liu, K. Jiang, H. Wang, C. Wang, X. Yang, F. Wu, Y. Leng, J. Liu, R. Li, Z. Xu, Nature, **595**(7868), 516 (2020).
11. J. D. Zuegel, S.W. Bahk, I. A. Begishev, J. Bromage, C. Dorrer, A. V. Okishev, and J. B. Oliver, CLEO 2014: Appl. and Tech. JTh4L-4 (2014).
12. A. Shaykin, I. Kostyukov, A. Sergeev, and E. Khazanov, Laser Review, **42**(2), 141 (2014).
13. E. Cartlidge, Science **355**(6327), 785 (2017).
14. F. Wu, J. Hu, X. Liu, Z. Zhang, P. Bai, X. Wang, Y. Zhao, X. Yang, Y. Xu, C. Wang, Y. Leng, R. Li, High Power Laser Sci. Eng. **10**, e38, (2022).
15. E. A. Khazanov, S. Yu. Mironov, G. Mourou, Physics-Uspekhi **62**(11), 1096 (2019).
16. T. Nagy, P. Simon, L. Veisz, Adv. Phys. X **6**, 1 (2020).
17. S. K. Lee, J. Y. Yoo, J. I. Kim, R. Bhushan, Y. G. Kim, J. W. Yoon, H. W. Lee, J. H. Sung, C. H. Nam, Ieee, in Conference on Lasers and Electro-Optics (CLEO), 2018.
18. V. Ginzburg, I. Yakovlev, A. Zuev, A. Korobeynikova, A. Kochetkov, A. Kuzmin, S. Mironov, A. Shaykin, I. Shaikin, E. Khazanov, G. Mourou, Phys. Rev. A **101,** 013829 (2020).
19. V. Ginzburg, I. Yakovlev, A. Kochetkov, A. Kuzmin, S. Mironov, I. Shaikin, A. Shaykin, E. Khazanov, Opt. Express **29**, 28297 (2021).
20. A. Shaykin, V. Ginzburg, I. Yakovlev, A. Kochetkov, A. Kuzmin, S. Mironov, I. Shaikin, S. Stukachev, V. Lozhkarev, A. Prokhorov, and E. Khazanov, High Power Laser Sci. Eng. **9**, e54 (2021).
21. J. I. Kim, Y. G. Kim, J. M. Yang, J. W. Yoon, J. H. Sung, S. K. Lee, and C. H. Nam, Opt. Express **30**, 8734 (2022).
22. A. A. Voronin, A. M. Zheltikov, T. Ditmire, B. Rus, and G. Korn, Opt. Commun. **291**, 299 (2013).
23. S. Mironov, V. Lozhkarev, G. Luchinin, A. Shaykin, and E. Khazanov, Appl. Phys. B **113**, 147 (2013).
24. M. Martyanov, V. Ginzburg, A. Balakin, S. Skobelev, D. Silin, A. Kochetkov, I. Yakovlev, A. Kuzmin, S. Mironov, I. Shaikin, S. Stukachev, A. Shaykin, E. Khazanov, A. Litvak, High Power Laser Sci. Eng. **11**, e28 (2023).
25. V.N. Ginzburg, I.V. Yakovlev, A.S. Zuev, A.P. Korobeynikova, A.A. Kochetkov, A.A. Kuz'min, S.Yu. Mironov, A.A. Shaykin, I.A. Shaikin, E.A. Khazanov, Quant. Electron. **49**(4), 29 (2019)
26. H. Zia, Communications in Nonlinear Science and Numerical Simulation **54**, 356 (2018).
27. X. Shen, S. Du, W. Liang, P. Wang, J. Liu, and R. Li, Appl. Phys. B **128**, 159 (2022).
28. J. Liu, X. Shen, S. Du, R. Li, Opt. Express **29**, 17140 (2021).
29. C. Wang, D. Wang, Y. Xu, and Y. Leng, Opt. Commun. **507**, 127613 (2022).